\documentclass[prl,twocolumn,amsmath,amssymb, a4paper]{revtex4}
\usepackage{natbib}
\usepackage{dcolumn}
\usepackage{graphicx}
\usepackage{bm}

\begin{document}

\newcommand{\degc}{\ensuremath{^{\circ}}C}

\bibliographystyle{apsrev}

\title{Three-dimensional Binary Superlattices of Oppositely-charged Colloids}
\author{Paul Bartlett\footnote[3]{To whom correspondence should be addressed (p.bartlett@bristol.ac.uk)}}
\author{Andrew I. Campbell\footnote[4]{Current address: Soft Condensed Matter Group, Debeye Institute, University of
Utrecht,  The Netherlands.}}

\affiliation{School of Chemistry, University of Bristol, Bristol BS8
1TS, UK.}

\date{April 29, 2005}


\begin{abstract}

We report the equilibrium self-assembly of binary crystals of
oppositely-charged colloidal microspheres at high density. By
varying the magnitude of the charge on near equal-sized spheres we
show that the structure of the binary crystal may be switched
between face-centered cubic, cesium chloride and sodium chloride. We
interpret these transformations in terms of a competition between
entropic and Coulombic forces.

\end{abstract}

\maketitle

The spontaneous crystallization of mixtures of large and small
colloids continues to fascinate material scientists since the first
observation of colloidal alloys, over two decades ago
\cite{Sanders-83}. There are several reasons for this continuing
interest: From a fundamental standpoint, colloids provide arguably
the best experimental realization of the hard sphere model, whose
phase diagram is completely dominated by entropy. Consequently the
rich variety of self-assembly phenomena seen in colloidal mixtures
provides a fascinating test bed for many-body statistical physics.
From a practical standpoint, colloidal assembly has great
technological potential as a fabrication technique for
three-dimensional photonic band gap (PBG) materials
\cite{3427,1642}, ultrafast optical switches \cite{3620} and
chemical sensors \cite{3623}. Most studies of colloidal assembly
have, to date, been restricted to the simplest situation, that of
purely repulsive spherical particles where the particles  bear
either no or similar surface charges and so behave as hard
spheres. Extensive computer simulations
  \cite{Eldridge-199,Trizac-1511} and experiments
 have located \cite{Bartlett-13,1685,2151,3007,3566} only three
equilibrium binary phases for hard spheres, $AB$, $AB_{2}$ and
$AB_{13}$, iso-structural with the atomic analogues NaCl, AlB$_{2}$
and NaZn$_{13}$.  The small number and the essentially close-packed
nature of these lattices limit the practical usefulness of colloidal
assembly techniques, particularly in the search for a visible PBG
material.  More complex architectures could, in principle, be
achieved by introducing weak reversible attractions between species
or by fine-tuning the inter-particle forces \cite{3596,3101,3100}.
Unfortunately all attempts to date to direct the assembly of
attractive binary particles have created highly-disordered
aggregates rather than the hoped-for crystalline structures.

 In this letter, we
demonstrate how the introduction of a weak reversible attraction
between unlike particles stabilizes  non-close packed
superlattices which would otherwise be entropically unfavorable. We
use fluorescent confocal microscopy to follow the self-organization
of a non-aqueous binary suspension of micron-sized spheres with
similar diameters but carrying opposite charges. The non-polar
environment used  ensures that the strength of the attraction is
comparable to $k_{B}T$ and all interactions remain reversible. We
study the phase behavior as a function of the relative strength of
the attractive and thermal interactions and find that we can tune
the translational crystal symmetry from face-centered cubic (fcc),
through body-centered cubic (bcc) to simple cubic (sc).

Our system consists of a binary mixture of near-identical sized
colloidal poly(methyl methacrylate) particles \cite{2320} suspended
in a non-polar density and index-matched solvent. Particles are
distinguished by containing one of two different cationic
fluorescent dyes. The large spheres (component $A$) have a core
radius of $R_{A}=777$ nm and contain the orange-red fluorescent dye
DiIC$_{18}$ (maximum emission at $\lambda_{em}= 565$ nm). The small
spheres (component $B$), with a core radius of $R_{B}=720$ nm, are
labeled with the green fluorescent dye DiOC$_{18}$ ($\lambda_{em}=
502$ nm). The radius ratio ($\gamma = R_{B}/R_{A}$) is accordingly
very close to unity ($\gamma = 0.93 \pm 0.01$).  Both particles have
a narrow distribution of sizes (polydispersity $\sigma$ of about
4\%) and individually form random hexagonal close-packed (rhcp)
crystals at high densities.  The volume fraction $\phi$ of the
samples was defined relative to the hard-sphere freezing transition
at $\phi_{f} = 0.494$. The solvent consists of a 78:22 wt \% density
and near-index matching mixture of cycloheptyl bromide (CHPB) and
\textit{cis}-decalin (CD). PMMA particles dispersed in this low
polarity solvent develop a small positive surface charge. The charge
on each particle is controlled by the polarity of the dye used as
the fluorescent label, with particles containing the  polar
DiIC$_{18}$ dye developing a larger charge than spheres labeled with
the less polar DiOC$_{18}$ molecule. The particle charge may be
reduced by increasing the Br$^{-}$ ion concentration in solution
\cite{3067}. Generating free Br$^{-}$ by irradiating the suspension
with ultraviolet light or by including a small length of
ferromagnetic wire to act as a  site for catalytic decomposition of
cycloheptyl bromide, led to a reduction in particle charge until at
the point of zero charge (PZC) the sign of the charge reversed and
then became progressively more negative with further increases in
[Br$^{-}$]. Charge inversion provides a useful, and to date
unexplored, tool to adjust the inter-particle interactions in a
binary suspension. By working near the point of zero charge the size
and sign of the charge on each species may be systematically
controlled and the total inter-particle potential ``tuned''. The
point of zero charge is a function of the initial particle charge
and so is different for DiIC$_{18}$ and DiOC$_{18}$-labeled
particles. At low (or high) Br$^{-}$ concentrations both particles
are positive (or negative), all interactions repulsive and a hard
sphere model is appropriate. However, between these two limits the
two spheres have opposite signs and the unlike interactions are
attractive. We found that, in our system, we could achieve partial
charge inversion by adding a small length of ferromagnetic wire to
our suspensions. The charge on the particle is a function of the
time, $t$, of contact between suspension and metal. At $t = 0$ both
particles are positively charged ($Z_{A} = +210$e, $Z_{B} = +100$e)
while after 48 hours the charge on the DiOC$_{18}$-labeled spheres
is reversed ($Z_{A} = +110$e, $Z_{B} = -40$e) and the two colloidal
species are oppositely charged. Using the measured charges and an
inverse screening length of $\kappa^{-1} = 400$ nm, as determined by
conductivity measurements, we estimate the repulsive potential
between unlike spheres at $t=0$ as $\sim +5k_{B}T$, at contact.
After 48 hours, charge inversion generates a cross-attraction which
is of order $\sim -3 k_{B}T$. These estimates highlight the weak and
reversible nature of the electrostatic interactions in our system.

Hard-sphere suspensions are fluid at low volume fractions but
crystallize into a random hexagonal close-packed structure at volume
fractions $\phi > 0.494$. Crystallization continues up to a volume
fraction $\phi_{g} \approx 0.58$ where a glass transition suppresses
further nucleation and growth. A binary mixture of hard spheres of
similar sizes is expected to be little different. For the size ratio
$\gamma = 0.93$ used here computer simulations \cite{Kranendonk-143}
and density functional calculations confirm that the two species
remain mixed in both the crystal and fluid phases and that there is
no phase separation. To explore the effect of the cross attraction
on the phase behavior we therefore prepared binary suspensions with
equal volumes of $A$ and $B$ particles and volume
fractions which ranged from $\phi = 0.48$, just below the freezing
transition up to $\phi = 0.58$, the expected glass transition. The
packing of the binary system of spheres was studied in real space
with two-color fluorescence confocal microscopy. The different
emission spectra of DiIC$_{18}$ and DiOC$_{18}$ dyes enable the two
types of particles to be readily discriminated. A stack of 200
images, spaced by 0.16 $\mu$m vertically, was collected from a 73
$\mu$m by 73 $\mu$m by 32 $\mu$m volume containing $\sim$ 20000
particles of each species. The suspension was sealed in a
cylindrical cell of $\sim 50$ $\mu$L volume containing a short $\sim
5$ mm length of ferromagnetic wire to catalyze the decomposition of
CHPB and which could be agitated with a magnet to rapidly mix the
sample prior to observation.

\begin{figure}[h] 
\includegraphics[width=3.0in]{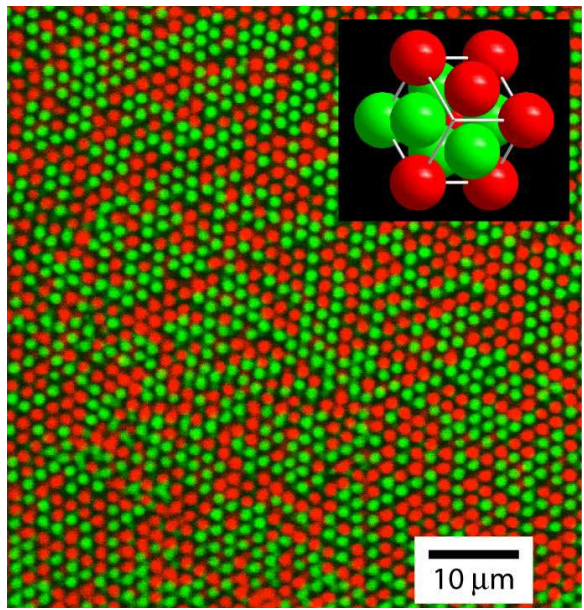}
\caption{Confocal image of substitutionally-disordered rhcp lattice
formed, after 23 hours, in a suspension with $\phi = 0.507$. The
image depicts the close-packed \{111\} plane. The larger spheres
(component $A$) are shown in green and the smaller spheres ($B$) in
red. The inset figure shows the \{111\} face of a fcc unit cell of
binary spheres with volume fraction $\phi_{c} = 0.55$.}
\label{fig:rhcp}
\end{figure}

Suspensions with total volume fractions $\phi \leq 0.487$ formed
amorphous fluids with strong short-range order but no long-range
correlations. An increase in volume fraction however led to the
rapid formation of binary colloidal crystals. We found three
distinct regular structures depending on the degree of charging, as
characterized by the contact time $t$. Weakly-charged particles at
high volume fractions assembled into a twelve-fold coordinated
substitutionally disordered random hexagonal close-packed crystal
(rhcp) (Fig.~\ref{fig:rhcp}). Computer simulations predict this to
be a stable phase in hard sphere mixtures of similar size
\cite{Kranendonk-143}. In this structure the spheres are located at
the crystal lattice sites in more or less random fashion. The lack
of strong correlations between the two species is clearly evident in
Fig.~\ref{fig:rhcp}.

 \begin{figure}[h] 
\includegraphics*[bb = 80 40 560 370, width=3.3in]{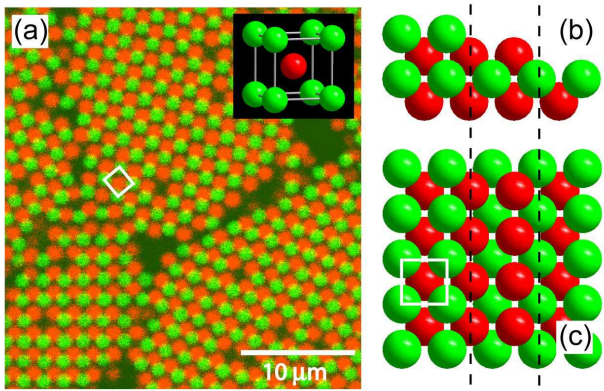}
\caption{Confocal image and model of cesium chloride superlattice
(SG 221) formed, after 199 hours, in suspension with $\phi = 0.528$.
(a) Confocal image of \{100\} plane in polycrystalline sample.
Colors as in Fig.~\ref{fig:rhcp}. Note the square lattice of $A$
spheres around each $B$. Inset shows CsCl unit cell. (b) and (c)
Depiction of a slice through a CsCl crystal, parallel to the \{100\}
plane, at three different depths of focus (dashed lines). (b)
Projection perpendicular to \{100\} plane. (c) In-plane projection
of \{100\} plane. Note the in-plane projection (c) remains unchanged
as the depth of focus is stepped down through the crystal. This
agrees with the experimental observation that the symmetry of the
confocal image (a) does not change as the depth of focus was
varied.} \label{fig:CsCl}
\end{figure}

Increasing the magnitude of the opposite charges on each particle
resulted in a completely different arrangement of large and small
spheres. Fig.~\ref{fig:CsCl}(a) shows a striking two-color confocal
micrograph of an AB superlattice formed in a suspension of $\phi =
0.528$ after 199 hours. Fig.~\ref{fig:CsCl}(b) depicts a
crystallographic model of the superlattice built from a cesium
chloride unit cell (space group SG 221). Rotation and cleavage along
the \{100\} plane reproduced the structures observed in the confocal
images. The cesium chloride structure consists of an equal number of
cesium and chloride ions arranged alternately at the vertices of a
body-centered cubic lattice so each particle has eight unlike
neighbors.   The lattice constant $a_{\rm{CsCl}} = 1.741 \pm 0.003$
$\mu$m,  determined from Fig.~\ref{fig:CsCl}, equates to a crystal
volume fraction of $\phi_{\rm{CsCl}} = 0.67 \pm 0.01$. The maximum
packing limit for a CsCl crystal formed from a mixture of
similarly-sized spheres ($\gamma > \sqrt{3}-1$) is
$\phi^{*}_{\rm{CsCl}}(\gamma) = \sqrt{3} \pi (1+\gamma^{3}) /
[2(1+\gamma)^3]$. For $\gamma = 0.93$ the corresponding maximum
volume fraction is $\phi^{*}_{\rm{CsCl}} = 0.68$, which indicates
that the CsCl superlattice observed  is dense packed and the
particles are close to touching each other. Note that although
dense, the CsCl crystal is not a maximally-packed structure. When
the two particles $A$ and $B$ are identically-sized, CsCl is
iso-structural with a bcc packing of spheres  which fills space
significantly less efficiently than fcc packing ($\phi_{\rm{fcc}} =
0.74$). Cesium chloride crystals were identified in binary samples
with a wide range of volume fractions $0.507 \leq \phi \leq 0.578$.
In each case CsCl growth always proceeded initial formation of a
substitutionally-disordered random-stacked close-packed crystal.

\begin{figure}[h] 
\includegraphics*[bb = 217 210 560 427, width=3.3in]{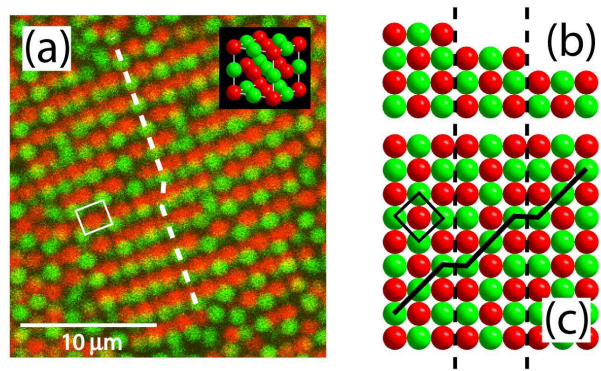}
\caption{Confocal image and model of sodium chloride superlattice
(SG 225) formed, after 215 hours, in suspension with $\phi = 0.528$.
(a) Confocal image of tilted \{100\} plane. Colors as in
Fig.~\ref{fig:rhcp}. Note the square lattice of $A$ spheres around
each $B$. Inset shows NaCl unit cell. (b) and (c) Depiction of a
slice through a NaCl crystal, parallel to the  \{100\} plane, at
three different depths of focus (dashed lines). (b) Projection
perpendicular to \{100\} plane. (c) In-plane projection of \{100\}
plane. Projection (b) reveals that the \{100\} plane consists of
vertical columns of alternate $A$ and $B$ spheres. Focusing down
through the \{100\} plane (c) causes a shift in the 2D
square-lattice of the face (shown in black) and a corresponding kink
in the line of spheres. The same characteristic pattern is evident
in the confocal image (white dashed line) confirming that the square
lattice seen in (a) arises from a NaCl rather than a CsCl structure.
} \label{fig:NaCl1}
\end{figure}

A further increase in the charge on the two particles led to the
formation of a third AB superlattice. Fig.~\ref{fig:NaCl1}(a)  shows
a  confocal micrograph of the superlattice formed 215 hours after
mixing in a binary mixture of total volume fraction $\phi = 0.528$.
Crystallographic modeling (Fig.~\ref{fig:NaCl1}(b)) reveals that the
ordered regions visible are consistent with cleavage along the
\{100\} and \{111\} (data not shown)  planes of a NaCl superlattice
(SG 225). The projection of the \{100\} plane visible in
Fig.~\ref{fig:NaCl1}(a) is superficially similar to the \{100\}
plane of the CsCl superlattice visible in Fig.~\ref{fig:CsCl}(a).
However the two structures are easily discriminated by stepping the
depth of focus. In the NaCl superlattice the vertically-aligned
columns of particles, visible in Fig.~\ref{fig:NaCl1}b, are composed
of alternating $A$ and $B$ particles in contrast to the CsCl
structure (Fig.~\ref{fig:CsCl}b). The NaCl structure is most
conveniently envisaged as two interpenetrating simple cubic
lattices, occupied by an equal number of either sodium or chloride
ions so that each ion has six unlike ions as its nearest neighbors.
Analysis gave an average unit cell length $a_{\rm{NaCl}}$ of 3.3
$\pm$ 0.2 $\mu$m which equates to a crystal volume fraction of
$\phi_{\rm{NaCl}} = 0.49 \pm 0.09$. The maximum packing fraction of
the NaCl superlattice is $\phi^{*}_{\rm{NaCl}}(\gamma) = 2\pi
(1+\gamma^{3}) / [3(1+\gamma)^3]$ for a mixture of spheres with a
size ratio $\gamma \geq \sqrt{2}-1$. For $\gamma = 0.93$ this
expression gives a maximum volume fraction of $\phi^{*}_{\rm{NaCl}}
= 0.53$, which is in reasonable agreement with the value measured.
The low packing fraction is readily understood if it is remembered
that a NaCl crystal is equivalent to sc packing when no distinction
is made between oppositely-charged spheres. It is striking that NaCl
is observed even though $\phi^{*}_{\rm{NaCl}}$ is considerably lower
than the maximum packing fractions of either CsCl
($\phi^{*}_{\rm{CsCl}} = 0.68$) or rhcp ($\phi^{*}_{\rm{rhcp}} =
0.74$).

The complex self-organization evident in our experiments is a
consequence of a subtle interplay between entropy which favors dense
close-packed structures and electrostatic interactions which prefer
more open non-close packed structures with reduced particle
coordination. The same balance plays a fundamental role in the study
of molten salts where it is frequently described in terms of the
restricted primitive model (RPM). The RPM consists of a mixture of
equal-sized hard spheres, half of which carry a positive charge and
the remainder an equal but opposite negative charge to ensure charge
neutrality. The ions $i$ and $j$ interact through a potential which
is a sum of a hard sphere repulsion and a purely Coulombic term,
\begin{equation}\label{eqn:rpm}
    u_{ij}(r) = \left \{ \begin{array}{ll}
    \infty & \textrm{if }r <
    \sigma \\
    \pm \; u_{c}  \left( \frac{\sigma}{r} \right) & \textrm{if }r \geq
    \sigma.
    \end{array} \right.
\end{equation}
In Eq.~\ref{eqn:rpm}, $u_{c}$ is the magnitude of the Coulombic
potential at contact, $\sigma$ is the hard sphere diameter, and the
plus and minus signs apply, respectively, to interactions between
ions of the same and different charge. The RPM is unlikely to be an
accurate model of our experiments as it does not incorporate
electrostatic screening. Nevertheless the RPM accounts at least
qualitatively for many of our observations.

The solid phases formed at high density in the restricted primitive
model have been investigated by a number of authors, both
theoretically \cite{3095,3119} and more recently by Monte Carlo
simulations \cite{3399,3115,3114}. The phase diagram is a function
of $\phi$ and $u_{c}/k_{B}T$, the relative strength of the Coulombic
potential at contact and the thermal energy. At low potentials
$u_{c} \alt 3.5 k_{B}T $,  where hard sphere repulsions dominate, a
fluid phase of ions is predicted to freeze at a density $\phi
\approx 0.5$ into a substitutionally-disordered close-packed fcc
crystal. The transition is driven by the higher entropy of the fcc
crystal as compared to a fluid of the same density.  Introducing
Coulombic interactions adds potential as well as entropic
contributions to the total free energy. In a purely random fcc
structure each ion is surrounded by six ions of the opposite charge
while in CsCl each ion has eight surrounding counterions.
Consequently we expect a fcc-CsCl transition as the Coulombic forces
are increased. Computer simulations shows that the CsCl crystal is
stabilized when $u_{c} \geq 4.3 k_{B}T$. At high densities and
charges the RPM predicts a further solid phase \cite{3115,3114}.
However the structure is expected to be tetragonal rather than the
cubic NaCl crystal observed here. Nonetheless it is well known that,
for a number of alkali halides, the free energies of the CsCl and
NaCl structures are sufficiently close that pressures of a few
hundred kilobar are sufficient to inter-convert the two structures.
Consequently our observation of NaCl may be a consequence of subtle
differences in the interaction potential from the purely Coulombic
mixture of oppositely-charged spheres assumed in the RPM.

 In summary we have demonstrated how the
introduction of a weak reversible attraction between equal-sized
particles leads to a surprisingly complex solid phase behavior with
observation of substitutionally-disordered rhcp, CsCl and NaCl
binary crystals. Although to date we have explored only a single
radius ratio ($\gamma = 0.93$) we have shown that a cross-attraction
between particles stabilizes several new non-close packed
structures, which are unstable in pure hard-sphere mixtures at the
same size ratio. Our study suggests that competition between entropy
and electrostatic interactions may provide a general route for the
synthesis of new and more complex binary structures than can be
stabilized by entropy alone and  highlights the potential to direct
the spontaneous growth of nano-particles into a diverse range of
superstructures by engineering the interactions between species.

We thank A. Donovan, K. Paul, V. Anderson and J. van Duijneveldt for
help with the experiments and analysis. The authors acknowledge
financial support from EPSRC, Grant No. GR/R17928/01.


\end{document}